\documentclass[twocolumn,numberedappendix,appendixfloats,iop,apj]{emulateapj}
\usepackage{mathrsfs}
\usepackage{graphicx}
\usepackage{amssymb}

\newcommand{\RG}{R_{\mathcal{G}}}

\newcommand{\Cov}{{{\rm Cov}}}
\newcommand{\Mpc}{{\ensuremath{\rm Mpc}}}

\newcommand{\km}{{\ensuremath{\rm km}}}
\newcommand{\K}{{\ensuremath{\rm K}}}
\newcommand{\mK}{{\ensuremath{\rm mK}}}

\newcommand{\veck}{\vec k}

\newcommand{\Rcv}{\mathcal{R}}
\newcommand{\ppri}{\prime\prime}
\newcommand{\pri}{\prime}
\newcommand{\genus}{\mathcal{G}}
\newcommand{\mA}{\mathcal{A}}
\newcommand{\mS}{\mathcal{S}}
     
\begin{document}

\title{Topology of large scale structure as test of modified gravity}

\author{
Xin Wang\altaffilmark{1,2},
Xuelei Chen\altaffilmark{1,3},
Changbom Park\altaffilmark{4}}

\altaffiltext{1}{Key Laboratory of Optical Astronomy, 
National Astronomical Observatories, Chinese Academy of Sciences,
Beijing, 100012, China}
\altaffiltext{2}{Graduate School of Chinese Academy of Sciences, Beijing 
100049, China}
\altaffiltext{3}{Center of High Energy Physics, Peking University, Beijing 
100871, China}
\altaffiltext{4}{Korea Institute for Advanced Study, Dongdaemun-gu, 
Seoul 130-722, Korea}


\begin{abstract}
The genus of the iso-density contours is a robust measure of the topology of 
large scale structure, and it is relatively insensitive to nonlinear 
gravitational evolution, 
galaxy bias and redshift-space distortion. 
We show that the growth of density fluctuations is scale-dependent even in the 
linear regime in some modified gravity theories, which opens a new possibility of 
testing the theories observationally. We propose to use the genus of the 
iso-density contours, an intrinsic measure of the topology of large-scale structure,
as a statistic to be used in such tests. In Einstein's general theory of relativity, 
density fluctuations are growing at the same rate on all scales in the linear 
regime, and the genus per comoving volume is almost conserved 
as structures are growing homologously, so we 
expect that the genus-smoothing scale relation is basically time-independent. However, in 
some modified gravity models where structures grow with different rates on different 
scales, the genus-smoothing scale relation should change over time. This can 
be used to test the gravity models with large scale structure observations. 
We studied the case 
of the $f(\Rcv)$ theory, DGP braneworld theory as well as the parameterized 
post-Friedmann (PPF) models. We also forecast how the modified gravity models 
can be constrained with optical/IR or redshifted 21cm radio surveys in the near future.
\end{abstract}


\maketitle

\section{Introduction}

The large scale structure has long been a major source of information on cosmic 
evolution. The most frequently used statistics of the large scale structure, such as 
the two-point correlation function and the density power spectrum, and especially 
the baryon acoustic oscillation (BAO) signatures in the power spectrum  
have been widely used in 
tests of cosmological models and precisional measurements 
of cosmological parameters \citep{T06,P07,G09}. 
However, these two-point statistics do not exhaust all of the information 
content about the large scale structure--at least not in the non-gaussian case. 
There are also some limitations on the application of these statistics. As 
the perturbations grow larger, their evolution becomes nonlinear, resulting in a
distorted power spectrum and the emergence of gravity-induced non-gaussianity.  
This has a crucial influence in the extraction of the absolute 
BAO scale: at $k=0.1 \sim 0.15 \Mpc/h$ where the first BAO peak is located,  
the non-linear 
evolution of the power spectrum amounts to 2\% to 5\% differences between $z=0$ and 
$z=0.5$ \citep{KP09}. Accurately modeling the non-linear evolution is 
highly non-trivial, and practical application in actual model test/parameter 
determination also requires fast and easy-to-implement algorithm.  
Although progresses have been made in more sophisticated perturbative 
expansion \citep{CS06a,CS06b,CS07,M08a,M08b,SS07} or reconstruction \citep{PW09,NW09} 
procedures to correct for the effect of non-linear evolution, a 
precise yet practical method to take into account of these effects has yet to be
developed. Other systematic effects, such as the clustering bias 
and the redshift distortion could also affect the outcome of the measurement. Such
effects may significantly limit the accuracy of cosmological tests 
for future redshift surveys, where the statistical error on the BAO scale would 
be at the one percent level.

An alternative approach to the correlation function and power spectrum is  
the topology of the iso-density contours, in particular the genus of these contours. 
This offers another way to characterize the statistical property of the large-scale
structure, and is insensitive to the systematic effects discussed above,
since the intrinsic topology does not change as the structures grow, at least not until
the iso-density contours eventually break at shell crossing\citep{PK09,Park05}. 
This provides a robust statistic for cosmology.
Compared with the power spectrum, the genus changes only by 0.5\% between 
$z=0.5$ and $z=0$ \citep{Park11}.
This is particularly true when using the volume fraction for the threshold levels
to identify the contours. According to the second order perturbation theory
\citep{M94}, there is no change in the genus at median density threshold due to
the weakly nonlinear evolution, because the iso-density contours enclosing 
a given fraction of volume do not change as long as the 
gravitational evolution conserves
the rank ordering of the density. Similarly, a monotonic clustering bias does not
result in any difference in the iso-density contours, and a continuous coordinate
mapping into redshift space does not affect the genus statistics either.
In $\Lambda$CDM model with general relativity, the genus is conserved 
during the different epochs in the linear regime of evolution.
Therefore, by observing the genus curve at the
different redshifts and smoothing scales, one can use it as a robust
standard ruler for cosmological measurements \citep{PK09}.

In the present paper, we consider the phenomenological consequences of 
alternative gravity theories.  
A number of such models, e.g. the $f(\Rcv)$ theory  
\citep{C04,C03,NO03,SHS07} and the DGP brane-world model \citep{DGP00}, 
have been proposed to explain the accelerated expansion of the Universe.
Unlike the general relativity, where the growth of the density fluctuation
is at the same rate on all scales during the linear evolution,
the modified large-scale gravitational forces induced by extra scalar 
field can in general introduce new scale dependence in the growth of 
structure, and therefore distort the genus curve in a time-dependent way.
This provides us a new tool to  distinguish GR from its various 
alternatives. It probes the combined effects of background expansion and 
the growth of the structures, and therefore is able to break the degeneracy
between the dark energy and modified gravity models from expansion data alone
\citep{L05,Z07}.
Furthermore, such a measurement also has the advantage that it is insensitive 
to nonlinear systematics. This is particularly useful because 
for some modified gravity models, such as the ones invoking the 
chameleon \citep{K04a,K04b} mechanism or the Vainshtein mechanisms \citep{V72,DDGV02}, 
it is non-trivial to treat the effect of nonlinear evolution on the 
power spectrum, and a conservative treatment on nonlinearity
may only yield fairly weak constraints on gravity models with even strong
signatures \citep{SPH07}.
The genus, on the other hand, is almost conserved 
during much of the evolution, thus allowing us to avoid dealing with such 
problems in the nonlinear regime. Instead, we may make the calculation in the
linear regime, and expect it to be almost unchanged until very late into the 
nonlinear regime.  

Below in Sec.II we briefly review the calculation of the genus density 
and the cosmological perturbation theory in the parametrised 
post-Friedmann framework. In Sec.III we discuss how the genus density
changes in the various modified gravity models. We forecast the 
prospects of observation with the future redshift surveys in Sec. IV 
and conclude in Sec. V. An estimate of statistical uncertainty of the 
genus measurement is given in the Appendix.

\section{Theory}

The genus density $\mathcal{G}$ is given by the genus $g$, i.e. 
(the number of holes - number of isolated regions) divided by the volume
$V$. Smoothing the tracer distribution by a Gaussian filter with scale $\RG$, 
then the iso-density contours of smoothed field can be identified
for a given variance-normalized density threshold $\nu \equiv \delta/\sigma(\RG)$ 
or volume fraction. 
The genus curve, i.e. the genus per unit volume
of the iso-density contours as a function of such threshold, could be measured.

In the particular case of Gaussian fluctuations, the genus curve can actually be calculated
analytically \citep{H86,D70}: 
\begin{eqnarray}
\mathcal{G}(\nu) = \frac{g}{V} (\nu) =  {\mathcal A} (1-\nu^2) e^{-\nu^2/2},
\end{eqnarray}
and the amplitude  ${\mathcal A}$ is related to the power spectrum by
\begin{equation}
\label{eqn:genus_A}
{\mathcal A} = \frac{1}{4\pi^2 ~3^{3/2} }\biggl ( 
\frac{ \int d^3k~ k^2 P(k) W(kR_{\mathcal{G}}) }
 {\int d^3k~ P(k) W(kR_{\mathcal{G}}) } \biggr)^{3/2}.
\end{equation}
where $W$ is the window function for smoothing.
The threshold level $\nu$ is related to the volume fraction $f$ on the high-density side
of the density contour surface via
\begin{eqnarray}
 f= \frac{1}{\sqrt{2\pi}} \int_{\nu}^{\infty} e^{-x^2/2} dx
\end{eqnarray}

Although the genus is related to the power spectrum in the Gaussian random field case, 
its measurement is completely independent of the power spectrum measurement, 
but rather by means of the Gauss-Bonnet theorem\citep{H86}
\begin{eqnarray}
    g = -\frac{1}{4\pi} \int \kappa~ d \mS ,
\end{eqnarray}
where $g$ is related to the integration of the Gauss curvature $\kappa = 1/(r_1 r_2)$  
over the surface, $r_1$ and $ r_2$ are the principal radii of curvature at 
the integration point. In practice, after smoothing the galaxies samples 
or simulation data into continuous density field,
one first makes a fine tessellation of the whole volume of the space with
small polyhedrons, e.g. cubes or truncated octahedrons, the iso-density surfaces 
can then be approximated by the polyhedral surfaces. 
Since the only nonzero contributions to the integrated 
curvature of the polyhedral surfaces are from the vertices and are 
equal to the angle deficits,
the genus of the iso-density contour at $\delta_c$ can be obtained by the summation 
of the angle deficit over all vertices \citep{GMD86,WGM87}. In this way, 
the genus of the iso-density contours is measured independent of the two-point statistics
such as the power spectrum or correlation function, though mathematically we 
know that they should be related by Eq.(\ref{eqn:genus_A}) in the Gaussian case.

As the structures grow, the fluctuations would become non-gaussian gradually, and the 
genus is no longer solely determined from the nonlinear 
power spectrum of the structure. However, the amplitude of the genus curve $\mA$
 does not change too much in the weakly non-linear regime,
because the genus is a topological indicator which is independent of simple 
growth of clustering without merger.
A general correction to $\genus(\nu)$ for the non-Gaussian field have been derived
up to the second order perturbation \citep{M94,M03} 
\begin{eqnarray}
&& \genus_{\rm 2nd}(\nu)= \mA 
e^{-\nu^2/2} \biggl \{ (1-\nu^2)\nonumber \\
&& -\sigma  \left[ \frac{S}{6}H_5(\nu) + \frac{3T}{2}H_3(\nu) + 3U H_1(\nu)  \right] 
 + \cdots  \biggr \} ,
\end{eqnarray}
where $H_n(\nu)=(-1)^n e^{\nu^2/2}(d/d\nu)^2 e^{-\nu^2/2}$ is the $n$-th order Hermit
polynomials,  $S$, $T$ and $U$ are the third order moments of $\delta$ and its 
gradient $\nabla \delta$.
At the median density threshold ($\nu=0$), these corrections are zero, so
the amplitude of the genus curve does not change. 
In practice, we calculate $\mA$ from a set of points measured between $\nu=-1$ and $+1$.
Since $H_{2n+1}(\nu)$ are all odd functions in $\nu$, $\mA$ would not be affected
by non-Gaussianity up to the second order.
At smaller scales, further investigation \citep{Park11} shows that the 
directly measured genus amplitude is better conserved 
than the shape of the power spectrum (Eq. \ref{eqn:genus_A}). 
The insensitivity of the genus to the nonlinear structure formation is not an 
artifact of the smoothing procedure, but reflects the more fundamental nature 
of topological property of the random field.

As demonstrated by \cite{PK09}, when the correct redshift-distance relationship 
$r(z)$ is adopted, one would be using the same comoving volume $V(z)= (D_A^2/H)(z)$ 
and smoothing scale 
$R_{\mathcal{G}}$ at different epochs. Here $D_A(z)$ is the 
angular diameter distance and $H(z)$ is the Hubble expansion rate.
Therefore the data at different redshifts actually enclose statistically 
almost the same amount of structures, and the amplitude of genus curve $\mathcal{A}$ 
would be almost the same at different redshifts. This can also be seen 
from Eq.~(\ref{eqn:genus_A}), which shows that $\mathcal{A}$ actually measure the 
slope of the power spectrum around the smoothing scale $R_{\mathcal{G}}$. 
Since in the linear regime of GR only the growth rate of the structure, not the shape of it, evolves, $\mathcal{A}$ would be conserved. 
However, if an incorrect $r(z)$ due to an incorrect 
cosmology is adopted, both $V(z)$ and $R_{\mathcal{G}}$ would be mis-estimated.
Since the topology of the structure is not scale-free, 
the genus enclosed in a wrongly
sized volume and smoothed with a wrong scale would
lead to deviation from the actual one, in this case
\begin{eqnarray}
 \mathcal{A} _Y(z, R_{\mathcal{G},Y}) R_{\mathcal{G},Y}^3 =  
\mathcal{A} _X(R_{\mathcal{G},X}) R_{\mathcal{G},X}^3,
\end{eqnarray}
where $Y$ is the adopted cosmology parameters while $X$ is the true 
cosmology. The smoothing scale $R_{\mathcal{G}}$ for different cosmologies 
are related to each other by 
\begin{equation}
(R_{\mathcal{G},X}/ R_{\mathcal{G},Y} )^3= 
(D_A^2/H)_X/(D_A^2/H)_Y ,
\end{equation} 
Utilizing this effect, topology of the large scale structure 
can serve as a standard ruler in cosmology.

Since the genus would be affected by modifying
the density perturbation through Eq.~(\ref{eqn:genus_A}),
we now consider the evolution of density perturbations in modified gravity. 
A wide range of modified gravity theories which satisfy the basic requirement
of being a metric theory where energy-momentum is covariantly conserved,
can be studied in the so called parametrized post-Friedmann (PPF) 
framework \citep{HS07}. For these theories, on superhorizon
scales structure evolution
must be compatible with background evolution, on intermediate scales
the theory behaves as a scalar-tensor theory with a modified Poisson equation, 
while on small scales, to pass stringent local tests, the additional scalar
degree of freedom must be suppressed. The evolution of linear perturbations in 
theories which satisfy these conditions can be characterized 
by a few parameters. 
With the Newtonian gauge temporal and spatial curvature scalar 
perturbation $\Psi$ and $\Phi$, we introduce 
\begin{eqnarray}
g \equiv \frac{\Phi+\Psi}{\Phi-\Psi}, \qquad \Phi_-\equiv \frac{\Phi-\Psi}{2}.
\end{eqnarray}
In the absence of anisotropic stress, $\Phi=-\Psi$, and $g=0$ for the GR case.
However, in modified gravity $g$ may not be zero. From causality 
considerations, on superhorizon scales the evolution
of metric perturbation must be determined entirely by the expansion rate and 
$g$ to the first order of $k_H$, with 
\begin{eqnarray}
\label{eqn:fzeta}
&
\lim_{k_H \to 0}  \biggl(  
\Phi^{\ppri} -\Psi^{\pri} -\frac{H^{\ppri}}{H^{\pri}}  - 
\bigl(\frac{H^{\pri}}{H} - \frac{H^{\ppri}}{H^{\pri}} \bigr) \Psi
\biggr) \nonumber \\ 
&
=  \mathcal{O}(k_H^2) = \frac{1}{3} f_{\zeta} k_H V_m.
\end{eqnarray}
where $' \equiv d/d\ln a$, $V_m \sim k_H $ is the velocity perturbation, 
$k_H=k/Ha$.  
However, to the next order, we can then introduce $f_{\zeta}$ to characterize 
the curvature perturbation on superhorizon scales.
The subscript $\zeta \equiv \Phi-V_m/k_H$, which is the curvature perturbation 
in matter comoving gauge, indicates the fact that the LHS of 
Eq.~(\ref{eqn:fzeta}) equals to $\zeta$ in the matter comoving gauge.
On subhorizon scales, the lensing potential satisfies the modified 
Poisson equation,

\begin{eqnarray}
\label{eqn:poisson}
k^2 \Phi_- = \frac{4\pi G}{1+ f_G} a^2 \rho_m \Delta_m (k),
\end{eqnarray} 
where $\Delta_m$ is the fractional density perturbation, 
$f_G$ parametrizes modification to Newton constant.
Equivalently, we can introduce a new scalar field ${\rm \varphi}$ which characterise
the extra source felt by nonrelativistic particles,
\begin{equation}
\label{eqn:poisson_psi}
-k^2 \Psi(k)= \frac{4\pi G}{1+f_G}a^2\rho_m\Delta_m (k)+ \frac{1}{2} k^2 
\varphi(k),
\end{equation}
where $\varphi(k)$ satisfies the equation
\begin{equation}
\label{eqn:phi}
 -k^2 \varphi (k) = \frac{8\pi G}{1+f_G} a^2 g(a,k)\rho_m \Delta_m(k).
\end{equation}
Thus in the subhorizon regime, 
the scale-dependent deviation from the GR is described by $\varphi(k)$. 
In summary, in the PPF parameterization,
the modified gravity model is characterized by $g(a, k)$, $f_G(a)$, $f_{\zeta}(a)$ 
and the superhorizon-subhorizon transition scale $c_{\Gamma}$. 
The evolution of these functions depends on the specific model. 
The fluctuation power spectrum for such 
theories can be calculated by using 
the public Boltzmann code ${\tt camb}$ with the PPF 
module \footnote{\rm http://camb.info/ppf/} \citep{F08a,F08b,LB02}.

\section{Models}

As a first example we consider the $f(\Rcv)$ theory,
\begin{eqnarray}
 S = \int d^4x \sqrt{-g} \bigl [
   \frac{\Rcv + f(\Rcv) }{16 \pi G} + \mathcal{L}_m
  \bigr],
\end{eqnarray}
where $\Rcv$ is the Ricci scalar and $\mathcal{L}_m$ is the matter 
Lagrangian density.
For a given expansion history $H(a)$, e.g. the effective dark energy equation
of state $w_e=-1$, in order to explain the late-time 
acceleration, the form of $f(\Rcv)$ can be determined from a second 
order differential equation
\begin{eqnarray}
- f_{\Rcv} (H H^{\prime} + H^2) + \frac{1}{6} f + 
H^2 f_{\Rcv\Rcv} \Rcv^{\prime} = 
 \frac{8\pi G}{3} \rho -  H^2  ,
\end{eqnarray}
where $\rho$ is the total energy density,
$f_{\Rcv}$ and $f_{\Rcv\Rcv}$ are the first and second derivatives of 
$f(\Rcv)$ with respect to $\Rcv$.
A given expansion history permits a family of 
$f(\Rcv)$ functions, the additional degree of freedom is usually 
characterized by $B_0$, the present day value of the function $B(a)$, which
is the square of the Compton scale given by
\begin{eqnarray}
B(a) \equiv \frac{f_{\Rcv\Rcv}}{1+ f_{\Rcv}} \Rcv^{\prime} \frac{H}{H^{\prime}}
\end{eqnarray}
$B_0$ is thus a model parameter of the $f(\Rcv)$ theory, which is to be 
constrained by structure growth.
Given $H(a)$ and $B(a)$, the metric ratio at superhorizon scale $g_{SH}$ can be 
obtained by solving the differential equation
\begin{eqnarray}
 \Phi^{\ppri} + \biggl ( 1- \frac{H^{\ppri}}{H^{\pri}} + \frac{B^{\pri}}{1-B}
+ B \frac{H^{\pri}}{H} \biggr ) \Phi^{\pri} \nonumber \\
+ \biggl ( \frac{H^{\pri}}{H} - \frac{H^{\ppri}}{H^{\pri}} + \frac{B^{\pri}}{1-B}
\biggr ) = 0,
\end{eqnarray}
and utilizing the relation $\Psi = (- \Phi - B \Phi^{\pri})/(1-B)$.
As shown by \cite{HS07}, the scale dependent $g(a,k)$ can be well fitted by 
the interpolation function
\begin{eqnarray}
\label{eqn:gak}
 g(a, k) =  \frac{g_{SH}+ g_{QS} (c_g k_H )^{n_g} }
{1+ (c_g k_H  )^{n_g}}.
\end{eqnarray}
where $g_{QS}=-1/3$, $c_g=0.71 B^{1/2}$ and $n_g=2$, with other PPF parameters
$f_{\zeta}=-1/3 g$, $f_G=f_{\Rcv}$ and $c_{\Gamma}=1$.

\begin{figure}[!htp]
\begin{center}
\includegraphics[width=0.45\textwidth]{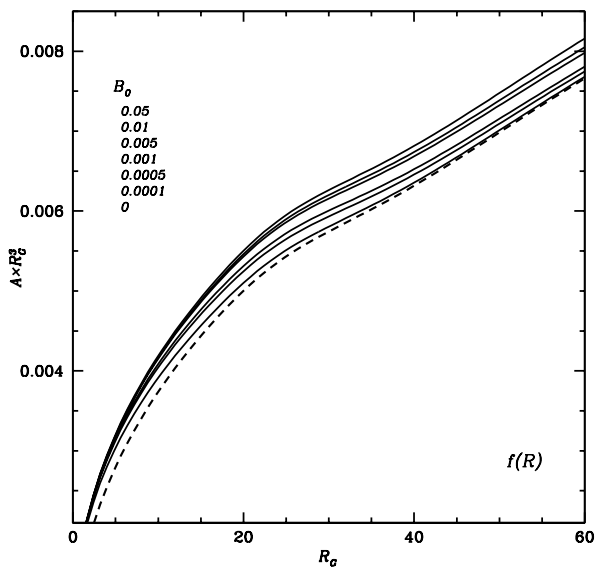}
\end{center}
\caption{\label{fig:fR_RG} The genus per smoothing volume 
as a function of smoothing scale $R_{\mathcal{G}}$ at $a=1$ for 
$f(\Rcv)$ theories with different $B_0$.
}
\end{figure}

Given PPF parameters of $f(\Rcv)$ theory above, especially $g(a,k)$, 
the matter power spectrum at relevant scales can be obtained
with the help of Eq.~(\ref{eqn:poisson_psi}, \ref{eqn:phi}).
In Fig.~(\ref{fig:fR_RG}), we plot the genus amplitude of the genus
as a function of the smoothing scale $R_{\mathcal{G}}$ at $a=1$
for several $f(\Rcv)$ models characterised by different $B_0$ values.
For models with greater $B_0$ value, the genus is larger.

\begin{figure}[!htbp]
\begin{center}
\includegraphics[width=0.45\textwidth]{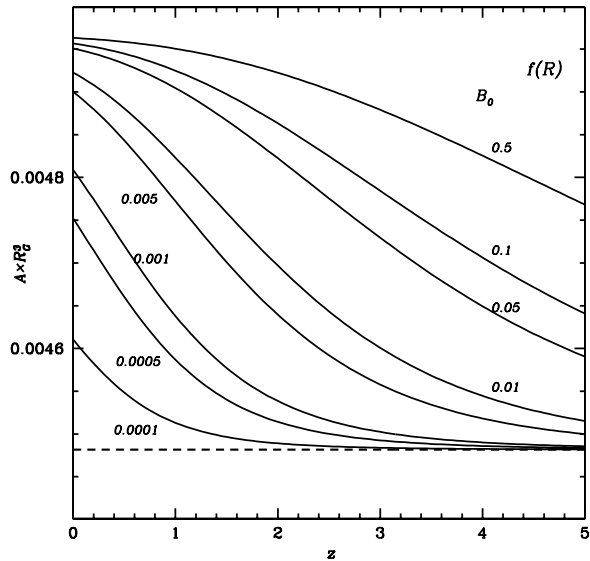}
\end{center}
\caption{\label{fig:fR_z} The redshift evolution of genus amplitude
for $f(\Rcv)$ theories with different $B_0$.
}
\end{figure}

We may also see how the amplitude of the genus curve varies as a function of
redshift. In Fig.~(\ref{fig:fR_z}), we plot
the redshift evolution of $\mathcal{A}$
for a fixed smoothing scale $R_{\mathcal{G}} = 15 \Mpc/h$.
Compared with the horizontal line, which corresponds to GR in the bottom 
of the figure, the amplitude of the genus curve exhibits a 
strong time dependence for the $f(\Rcv)$ theory with a total variation of
about $10\%$ for $B_0 \gtrsim 0.01$. 

From these figures it is obvious that for larger $B_0$, the deviation 
from GR is greater, i.e. the variation of the genus 
is stronger in the $f(\Rcv)$ theory.
By inspecting the perturbative variables in detail, with other PPF parameters
we find that this is mainly due to the monotonically increasing deviation of
$\Phi_-$ from the GR prediction with respect to the wavenumber $k$.
The extra scalar degree of freedom $\varphi=-f_\Rcv$ 
enhances the gravitational force at smaller scales, and therefore 
increase the slope of the clustering.

Next we consider the self-accelerating Dvali-Gabadadze-Porrati (DGP) braneworld 
model. In this model, our universe is a (3+1) dimensional brane embedded
in an infinite Minkowski bulk, it can be described by the following 
action
\begin{equation}
 S =  \int d^5x \sqrt{-g_{(5)}} \biggl [
\frac{\Rcv_{(5)} }{16 \pi G_{(5)}} + 
\delta(\chi) 
   \biggl( \frac{\Rcv_{(4)} }{16 \pi G_{(4)}}  + 
  \mathcal{L}_m \biggr) \biggr],
\end{equation}
where $\Rcv_{(i)}, G_{(i)}$, $i=4,5$ is the Ricci scalar and Newton constant
in the brane and the bulk respectively.
At the background level, \cite{DGP00} showed that the consequent Friedmann 
equation
\begin{eqnarray}
 H^2(a) =  \biggl ( \sqrt{ \frac{8\pi G_{(4)}}{3}  \sum_i \rho_i
 + \frac{1}{4 r_c^2} } + \frac{1}{2 r_c}
 \biggr)^2 - \frac{K}{a^2},
\end{eqnarray}
leads to a de-Sitter phase at late time characterized by the crossover scale
$r_c= G_{(5)}/2G_{(4)}$.

Above the horizon, the evolution of the perturbation can be solved by 
iterative scaling method \citep{S07}, and the metric ratio $g_{SH}$ is
well fitted by function
\begin{eqnarray}
g_{SH}(a) = \frac{9}{8 H r_c -1} \biggl( 1+ \frac{0.51}{H r_c- 1.08}\biggr).
\end{eqnarray}
In the quasi-static regime,
\begin{eqnarray}
g_{QS}(a) = -\frac{1}{3} \biggl[ 1- 2H r_c \biggl( 1+ \frac{1}{3} 
\frac{H^{\pri}}{H} \biggr) \biggr]^{-1}.
\end{eqnarray}
Thus the scale dependent $g(a,k)$ can be described by the same interpolation 
equation Eq.~(\ref{eqn:gak}) with $c_g=0.4, n_g=3$. Other PPF parameters are
$f_{\zeta}=0.4 g, f_G=0, c_{\Gamma}=1.$

From the calculation, we find that the major scale-dependent modification 
to the growth of the structure is in the superhorizon regime ($k/aH \ll 1$). 
With a Gaussian smoothing scale of $15 \Mpc/h$, 
only about $0.07\%$ deviation of $\mathcal{A}$ are found
compared with GR, which can hardly be distinguished by the observation in 
the near future.
In this case other techniques such as the ISW effect 
or the growth rate are more viable.
However it should also be noticed that \citep{S09,CS09}
with the help of Vainshtein mechanism which bring the gravity
back into GR at small scales, the nonlinear evolution can also induce 
significant scale-dependent deviations at relevant scales. 
We leave this topic to our further study.

At last we consider more generic models in the context of PPF, this 
allows the method to be applied to other possible modified gravity models,
and also the analysis would help us to see for whick kind of model 
the topological method is more sensitive.
We note that amongst the PPF variables, $f_G(a)$ and $f_{\zeta}(a)$
connect the metric fluctuations to the matter fields and are only 
functions of time.
Rescaling the amplitude of the matter fluctuations without 
changing the shape (i.e. slope) of the power spectrum 
would not affect the genus, so here we do not need to consider them, as the
topological measurements would be insensitive to these. Similarly, 
for the wavenumber range corresponding to smoothing scale of the order
$\mathcal{O}(10) \Mpc/h$ relevant to large-scale genus topology study, 
the mode is in the sub-horizon regime, 
so the variable $g_{SH}$ which describes the modification to the 
superhorizon metric can also be excluded from our consideration.

Following \cite{H08}, we adopt the scale dependent $g$ as an 
interpolation function between the super-horizon regime $g_{SH}$ and
the quasi-static regime $g_{QS}$ (c.f. Eq.~(\ref{eqn:gak}).
Both $c_g(a)$ and the power index $n_g$ can affect the behavior of 
the transition between the two regimes.

\begin{figure}[!htbp]
\begin{center}
\includegraphics[width=0.45\textwidth]{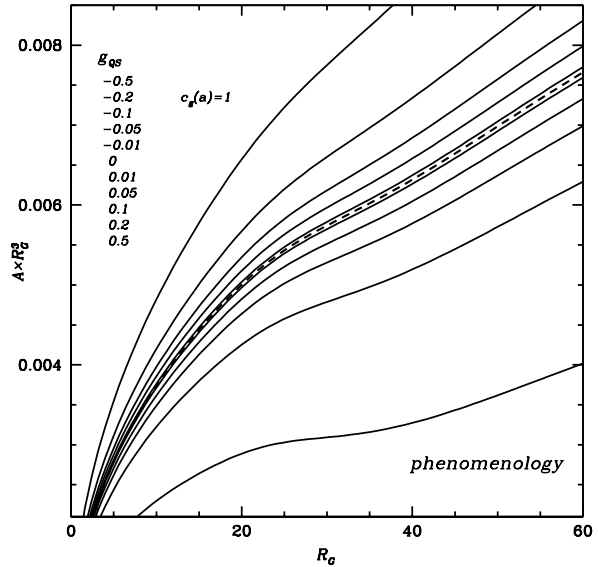}
\caption{\label{fig:phen_RG} The genus per smoothing volume 
as a function of smoothing scale $R_{\mathcal{G}}$ at $a=1$ for the 
phenomenological model of different $g_{QS}$ with $c_g(a)=1$.}
\end{center}
\end{figure}

Eq.~(\ref{eqn:gak}) shows that
for each Fourier  mode the time-dependent factor $c_g k_H= c_g k/(aH)$ 
determines the moment and scale at which the gravity deviate 
from the GR case.  In this simple model, 
 $\lim_{k \to 0} g(a,k) = 0$, therefore $\varphi$ would be modified when
$ k \geqslant k_t=(aH)/c_g$.

First let us consider the case where $c_g$ is a constant.
The transition scale would first slowly decrease as 
$k_t \sim a^{-1/2}$ during the matter dominated era, 
then start to increase during the era of accelerated expansion. 
In order to get significant effect on scales of 
$k \sim \mathcal{O}(0.1)h/\Mpc$ or above, 
$c_g$ should be at least larger than 
 $ (aH)/k $, which is about $\mathcal{O}(0.01)$.

As an example, 
in Fig.~(\ref{fig:phen_RG}), we plot the genus per smoothing 
volume as a function of $R_{\mathcal{G}}$ for various $g_{QS}$, while keeping
$g_{SH}=0$, $c_g(a)=1$, and $n_g=2$. 
Compared with the GR case (i.e. $g_{QS}=0$), the difference is quite apparent. 
At any time, the metric deviation Eq.~(\ref{eqn:gak}) is a monotonic function
of the wavenumber, this induce the same monotonicity in the deviation of 
$\Phi_-$ from GR.
Specifically, negative value of $g_{QS}$ and therefore the positive extra 
force $ -\nabla \varphi$, 
will increase $\Phi_-$ as well as the 
slope of power spectrum at $k\gtrsim k_t$, and ultimately 
rise the value of the genus.
On the other hand, positive $g_{QS}$ behaves in the opposite way.
Nevertheless, owing to the slow evolution of $k_t$, the variation of 
$\mathcal{A}$ between the present and $z=5$ is less than $1\%$ for 
$|g_{QS}|<0.5$.

\begin{figure}[ht]
\begin{center}
\includegraphics[height=0.45\textwidth]{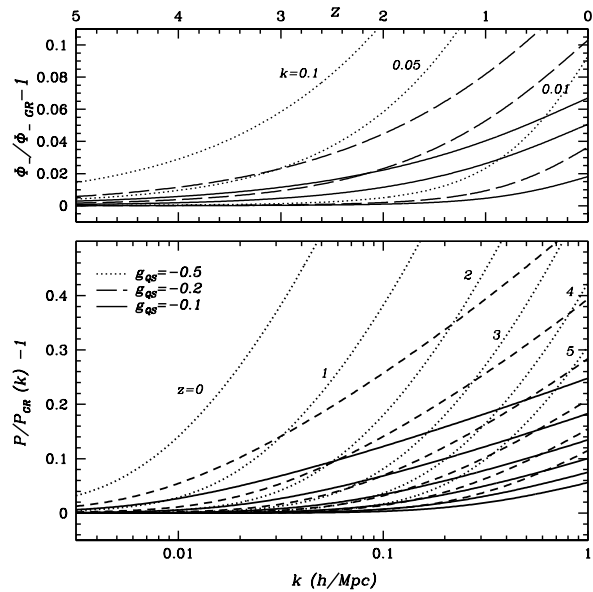}
\caption{\label{fig:power_metric} Deviation of $\Phi_-$ (upper) and 
power spectrum $P(k)$ (lower) compared with GR, assuming $g(a,k)$ has the 
form of Eq.~(\ref{eqn:gak}) and Eq.~(\ref{eqn:cg}), $c_g(0)=0.1$.}
\end{center}
\end{figure}

Next, we consider models in which $c_g$ varies with time. Here 
the deviation scale $k_t$ of the theory can vary significantly with time. 
As a toy model, let us assume that $c_g(a)$ depends on 
the square of the scale factor 
\begin{eqnarray}
\label{eqn:cg}
  c_g(a) = c_g(0) \cdot a^2, 
\end{eqnarray}
and parametrize the model by the value of $c_g(0)$. In this case, 
$k_t$ will always decrease,
and for a fixed wavenumber the absolute value of the deviation $|g|$ 
will always increase with time.  
Consequently, as illustrated in 
Fig.~(\ref{fig:power_metric}), $\Phi_-(k)$ (upper panel) 
and the power spectrum (lower panel) will deviate from the GR case 
progressively.

\begin{figure}[!htbp]
\begin{center}
\includegraphics[width=0.44\textwidth]{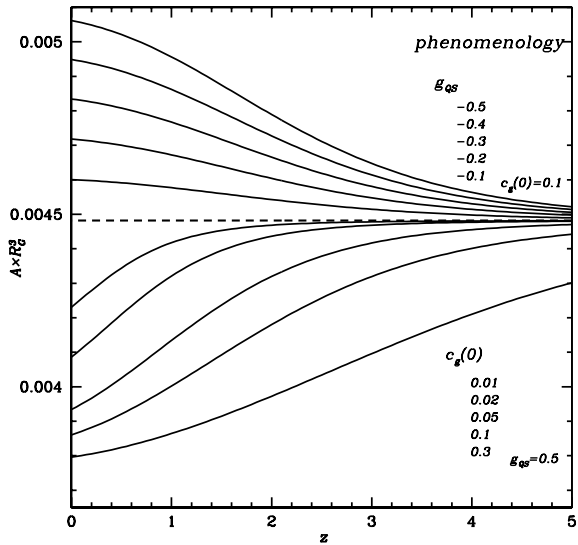}
\caption{\label{fig:phen_z} The redshift evolution of genus amplitude
for phenomenological models described by Eq.(\ref{eqn:gak}) and 
Eq.(\ref{eqn:cg}). The upper panel shows models with different $g_{QS}$
while keeping $c_g(0)=0.1$. The lower panel shows models with different 
different $c_g(0)$ while keeping $g_{QS}=0.5$.
The dashed line corresponds to the GR case in each panel.
}
\end{center}
\end{figure}

In Fig.~(\ref{fig:phen_z}),
we can see the evolution of the genus for different 
model parameters. In the upper panel, $g_{QS}$ is varied, 
taking values from -$0.5$ to -$0.1$ while $c_g=0.1~ a^2$. 
As expected, for the cases with larger value of $g_{QS}$,
the genus amplitude take a faster raise. In the lower panel, 
We show the result of varying the value of $c_g(0)$, while keeping  
$g_{QS}=0.5$ fixed.
In this case, increase $c_g(0)$ would move $k_t$ to larger scales, 
and shift the onset of the deviation to an earlier epoch,
so the genus should deviate from GR further and earlier. 
However, when $c_g(0) \gtrsim \mathcal{O}(1)$, the modification would move to
superhorizon regime, then it would become saturated with a nearly
constant genus.

\section{Prospects of Observation}

The genus curve of iso-density contours could be measured with future 
large scale structure surveys. In the following, we consider how 
such surveys could be used to constrain the modified gravity models with 
the topological measurements discussed above.

\begin{table*}[ht]
\begin{center}
\caption{\label{tab:survey} Parameters of future large scale
structure surveys.}
\begin{tabular}{ccccccc}
\hline
  Galaxies Survey & & redshift range &  &  sky area & &\\
\hline
BOSS    & & $ 0.2<z<0.7 $, &  &  $10000 ~ \deg^2$ &  &  \\     
LAMOST  & & $ 0.0 <z<0.7 $,&  &  $8000 ~ \deg^2$  &  &  \\      
WFMOS   & & $0.5<z<1.3$,   &   &  $2000 ~\deg^2$  &   &  \\      
JDEM    & & $0.5<z<2.0$,   &   &  $10000 ~\deg^2$ &   &  \\      
BIGBOSS & & $ 0.2<z<2.0 $, & &  $24000 ~ \deg^2$  & & \\     
\end{tabular}

\begin{tabular}{cccc}
\hline
  21cm Intensity Mapping& redshift range & other parameter & integration time (hour)\\
\hline
MWA     &  $3.5<z<5$, & $N_a=500 $, $A_s=16\pi^2~\deg^2$   & 1000\\
MWA5000 &  $3.5<z<5$, & $N_a=5000$, $A_s=16\pi^2~\deg^2$   & 4000\\
CRT  &  $0<z<2.5$, & $L=100m$, $W=15m\times7$,  &   10000\\
\hline
\end{tabular}
\end{center}
\end{table*}

The variance of the actual genus measurement is usually estimated with 
the help of simulation. For the purpose of making forecast, an analytical 
calculation is more desirable. In principle, the uncertainty in the 
topological measurement is different from the uncertainty in the power
spectrum of the large scale structure, this is especially true for 
the general (non-Gaussian) case. However, as discussed previously, the Gaussian
assumption should be a reasonable approximation for the purpose of making forecast,
as far as the large scales are concerned. 
In the appendix, we estimate the minimal amount of variance
$\sigma_\mathcal{A}$ (Eq. \ref{eqn:sigA}) by propagating from the uncertainty in
the power spectrum $\sigma_P(k)$. 
We also compared our analytical estimation with the measurement from \cite{GC09}, 
who utilized two volume-limited subsamples of luminous red galaxies (LRG) 
to measure the genus statistics: a dense shallow sample at $0.2<z<0.36$ with 
smoothing length $R_{\genus}=21 h^{-1}/\Mpc$, and a sparse deep sample at $0.2<z<0.44$
with $R_{\genus}= 34 h^{-1}/\Mpc$. Their result is consistent with Gaussian
distribution with the amplitude $A=167.4\pm 7.0$ and $A=79.6\pm 6.0$ 
respectively. By assuming a reasonable bias factor $b\sim 2$, our formula
Eq.(\ref{eqn:sigA}) gives $\sigma_A/A \sim 4.5\%$ for the shallow sample and $\sigma_A/A
\sim 7.4\% $ for the deep sample, which are very close to the measured values
($4.1\%$ and $7.5\%$ respectively). This shows that  
Eq.(\ref{eqn:sigA}) could indeed give good analytical estimate of $\sigma_A$.

Throughout this section, we assume the Gaussian smoothing scale 
$R_{\mathcal{G}}=15\Mpc/h$ unless explicitly emphasized otherwise.
At high redshifts the non-linear scale decreases, one can use smaller
$R_{\mathcal{G}}$ which makes the uncertainty of the genus much smaller.

Taking $\mathcal{O}=\ln\mathcal{A} $ as the observable, 
the Fisher matrix is 
\begin{equation}
\label{eqn:fisher}
 F_{ij} = \sum_{ \rm redshift ~bins } \frac{\partial \mathcal{O}}
{\partial p_i} ~~ \frac{1}{\sigma^2_{\mathcal{O}}}  ~~
 \frac{\partial \mathcal{O}} {\partial p_j} + F_{{\rm CMB}, ~ij},
\end{equation}
where  $p_i$ include 
$\Omega_m h^2, \Omega_b h^2, h_0, n_s, \ln A_s, \tau , w_e  $,
as well as other modified gravity parameters, e.g. $B_0$ for $f(\Rcv)$ theory. 
The Planck prior is added effectively by the contribution $F_{\rm CMB}$, which 
helps to break the degeneracies between the various parameters.
The summation is over different redshift bins with size of $\Delta z = 0.1$.
For cosmological parameters ($ \Omega_m h^2, h_0, w_e $),  both the 
perturbations and background evolution contribute to the constraints.
Therefore, derivatives in Eq.(\ref{eqn:fisher}) should be carefully calculated 
by taking into account the scaling relation 
\begin{eqnarray}
\mathcal{A}_{\rm obs}(R_{\mathcal{G}}) & = &
  \lambda ~ \mathcal{A}_{\rm true}
   (\lambda^{1/3} R_{\mathcal{G}} ) \\
\lambda(z) & = &\frac{ D_{\rm A,ref}^2 H}{ D_{\rm A}^2 H_{\rm ref}}.
\end{eqnarray}
Here $D_{\rm A, ref}$ and $H_{\rm ref}$ are the angular diameter distance
and the Hubble expansion rate evaluated in the reference cosmology, 
which is used when reconstructing the position from redshift, and
for simplicity 
we assumed it as the same as the fiducial cosmology.

\begin{figure}[ht]
\begin{center}
\includegraphics[height=0.45\textwidth]{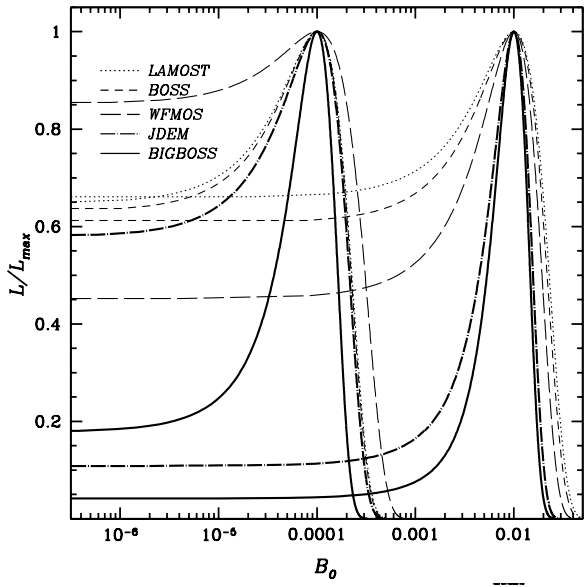}
\caption{\label{fig:cont_fr} Likelihood distribution on $B_0$ of 
$f(\Rcv)$ theory for various optical/IR surveys. Two fiducial value
($B_0=0.01$ and $0.0001$) are shown in the figure.}
\end{center}
\end{figure}

We consider the measurement of large scale structure topology in extended redshift
ranges with a number of optical/near IR galaxy surveys at $z<2$, such as the 
LAMOST\footnote{{\it http://www.lamost.org/}} (see also \cite{W09}), 
BOSS\footnote{{\it http://cosmology.lbl.gov/BOSS/, http://www.sdss3.org/ }}, 
WFMOS\citep{BNE05}, JDEM\footnote{{\it http://www.jdem.gsfc.nasa.gov/}}
and BIGBOSS\citep{SB09} survey, as well as a few 
21cm intensity mapping experiments at $0<z<5$,  e.g. the CRT\citep{C07,SS09} 
and the MWA\footnote{{\it http://www.MWAtelescope.org/}}. The survey
parameters we adopt  are listed in  Table~(\ref{tab:survey}). Needless to say,  
these parameters are only prelimnary estimates based on current planning, 
the parameter for the actual projects are subject to change.

In Fig.~(\ref{fig:cont_fr}), we illustrate the likelihood distribution over
$B_0$ of the $f(\Rcv)$ theory at fiducial values $B_0=10^{-2}$ and $10^{-4}$
calculated for various galaxies surveys. 
As one of the projects under planning, the BIGBOSS can 
provide the most stringent constraints, the 1-$\sigma$ error on $B_0$ is
$0.0039$ at $B_0=10^{-2}$ and $5.38\times 10^{-5}$ at $B_0=10^{-4}$.
Even for surveys which are on going or will start in the near future (e.g.
BOSS and LAMOST), one could also gain considerably rich information about
the redshift evolution of the genus amplitude. 
The 1-$\sigma$ errors at the same fiducial values are $0.011$ and 
$1.1\times10^{-4}$ for LAMOST,
$0.010$ and $1.04\times10^{-4}$ for BOSS respectively.
Notably, the constraining conclusion drawn from particular survey depends on 
the true cosmology, not only quantitatively but also qualitatively. 
For example, the WFMOS survey, which is deep but narrow and is more
powerful than BOSS and LAMOST survey at $B_0=10^{-2}$, become insignificant at
$B_0=10^{-4}$.  This is because when $B_0$ is small, 
the amplitude of genus at high redshift would quickly approach to 
the value of GR (c.f. Fig.~\ref{fig:fR_z}) and become nearly indistinguishable 
for a narrow survey.

\begin{figure}[htbp]
\begin{center}
\includegraphics[height=0.45\textwidth]{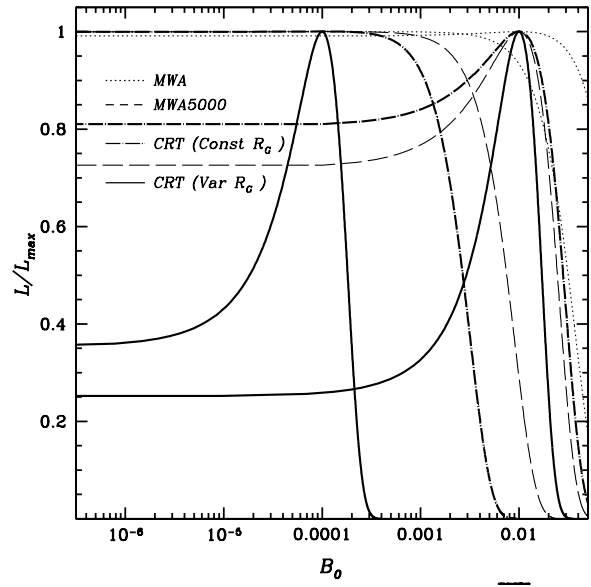}
\caption{\label{fig:cont_fr_21cm} The same plot as Fig.(\ref{fig:cont_fr}) 
for 21cm surveys.}
\end{center}
\end{figure}

Unlike galaxies surveys, the 21cm intensity mapping experiments have relatively
poor angular resolutions, for the CRT
\begin{eqnarray}
  R_{\rm res}(z) = r(z) \frac{\lambda}{L_{\rm CRT} }
\end{eqnarray}
can reach to the order of $10 \sim 30 \Mpc/h$ at higher redshift.
Here, $r(z)$ is the comoving angular diameter 
distance at redshift $z$, $\lambda$
is the received wavelength and $L_{\rm CRT}$ is the length of the cylinder of 
the CRT telescope.
Therefore the smoothing scale $R_{\mathcal{G}}$ when measuring the genus
should be larger than $R_{\rm res}$.

We plot in Fig.~(\ref{fig:cont_fr_21cm}) the likelihood of $B_0$
for the 21cm intensity mapping. For the CRT,
two different choices of $R_{\mathcal{G}}$ are considered here.
One is the constant $R_{\mathcal{G}}$ model  by assuming $R_{\mathcal{G}}= 2\times
\max \{ R_{\rm res}(z) \}$, which equals to $60\Mpc/h$ in this case.
The figure shows that the constraint of this model is not very
stringent for the $f(\Rcv)$ model.
Another choice is to smooth the data with varying $R_{\mathcal{G}}$ 
at different redshifts.
We divide the whole redshift range ($0<z<2.5$) into 4 
intervals, and assume the same factor of two relationship between the smoothing 
scale and the angular resolution within each redshift interval. In this case,
the 1-$\sigma$ error equals to $0.0060$ at $B_0=10^{-2}$ 
and $6.93\times10^{-5}$ at $B_0=10^{-4}$.
On the other hand, the angular resolution of the MWA and 
MWA5000 is sufficiently good so that we assume the same smoothing scale as
the optical/IR case. However, for their high redshift coverage,
the constraints are still not very stringent, especially at $B_0=10^{-4}$.

Numerous efforts have been made to constrain the $f(\Rcv)$ model,
with various observations such as the CMB anisotropies, supernovae, 
BAO distance, weak gravitational lensing, galaxies flow and clusters abundance. 
The currently strongest constraints combining all of the data gives 
$B_0<1.1\times 10^{-3}$ at $95\%$ C.L.\citep{LS10}. The main constraining 
power comes from the low redshift cluster abundance data, while the 
integrated Sachs-Wolfe (ISW) effects from galaxies-CMB cross correlation 
also provides a moderate constraint of   
$B_0<0.42$ at $95\%$ C.L.\citep{LS10}.
On the other hand,  the galaxies power spectrum data\citep{T06}
together with CMB\citep{WMAP06} and supernovae\citep{A05}, only places 
an upper bound on $B_0$ of order unity \citep{SPH07}, 
Although theoretical calculation shows significantly 
enhanced growth of large scale
structure in the $f(\Rcv)$ models\citep{SHS07}, uncertainties in the 
nuisance parameters such as the
galaxies bias $b$ and nonliearity parameter $Q_{nl}$ substantially weakened  
its constraining power. 
Assuming similar survey parameter as those in \cite{SPH07}, we find
the 1-$\sigma$ uncertainty $\sigma_{B_0}$ from genus measurement 
is around $0.2$  at $B_0 = 10^{-2}$. This demonstrates how the 
topological measurement which is insensitive to nonlinearity and clustering
bias could improve the large scale structure constraint on the $f(\Rcv)$ model,
though we should note that in real measurement the error might be 
larger than statistical estimates given here.


\begin{figure}[ht]
\begin{center}
\includegraphics[height=0.45\textwidth]{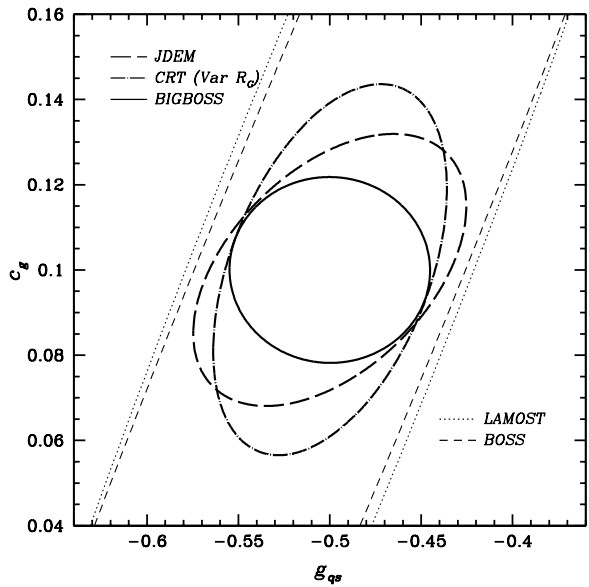}
\caption{\label{fig:cont_metric} 1-$\sigma$ constraint on $g_{QS}$
and $c_g(0)$, assuming $c_g (a)=c_g(0) \cdot a^2$}
\end{center}
\end{figure}

In Fig.~(\ref{fig:cont_metric}), we also plot the 2D constraint on 
$g_{QS}$-$c_g(0)$ for the phenomenological model considered in 
Eq.~(\ref{eqn:gak},\ref{eqn:cg}). 
Similar to the case of $f(\Rcv)$ model, the BIGBOSS provides a very
 stringent constraint, the 1-$\sigma$ errors are 
$\sigma_{g_{QS}}=0.036$ and $\sigma_{c_g(0)}=0.014$ respectively.
For the CRT, $\sigma_{g_{QS}}=0.042$ and $\sigma_{c_g(0)}=0.028$ 
(assume the redshift-varying $R_{\mathcal{G}}$).
$\sigma_{g_{QS}}=0.18$ and $\sigma_{c_g(0)}=0.19$ for the LAMOST,
and $\sigma_{g_{QS}}=0.17$ and $\sigma_{c_g(0)}=0.18$ for the BOSS.

\section{Conclusion}

The topological indicators such as the genus of the isodensity contours 
provides an independent way for characterizing 
the large scale structure, complementary to  
the more often used two point statistics such as the 
correlation function and power spectrum. A significant advantage of the   
the topological measurement is that it is less susceptible to 
the non-linear evolution and bias of the large scale structure, 
thus reducing the effects of these potential systematic errors, and may therefore
be an more reliable way to extract the information on the large scale structure.

In this paper, we studied the topology of the large scale structure 
as a measure of the scale-dependent expansion history of the universe
in models of modified gravity theories.
In the modified gravity theory models, the structure growth can be 
scale-dependent even in the linear regime, 
the amplitude of the genus curve varies significantly 
with redshift, and hence it can be
used as a new tool for distinguishing the models from the case of 
standard gravity (i.e. GR). 
We illustrated this for the $f(\Rcv)$ theory, DGP braneworld model as well 
as a phenomenological model parametrized with the PPF variables introduced by 
\cite{HS07}. We find that the genus curves for these models are modified 
and evolve with redshift due to the scale-dependent growth effect, hence the 
genus curve can be used as an observable to distinguish the modified gravity 
models from the general relativity theory. Finally, using the Fisher matrix formalism, 
we also made forecast on the sensitivity of this test with some 
current or future optical/IR and 21cm redshift surveys, showing that the method
is a competitive way to test modified gravity models.

\section*{Acknowledgements}

This work is supported by
the National Science Foundation of China under the Distinguished Young
Scholar Grant 10525314; by the
Chinese Academy of Sciences under grant KJCX3-SYW-N2; by the
Ministry of Science and Technology of China under the National Basic Science
program (project 973) grant 2007CB815401; and by 
the National Research Foundation of Korea (NRF) grant
funded by the Korea government MEST (No. 2009-0062868)

\section*{Appendix: Statistical Uncertainty of the Genus Measurement}

Since the genus amplitude $\mathcal{A}$ can be expressed as a function
of the power spectrum (Eq.~\ref{eqn:genus_A}), we may approximately estimate 
the statistical uncertainty $\sigma_{\mathcal{A}}$ by propagating the error
from the uncertainty of power spectrum $\sigma_P(k)$. 

Rewriting the observable 
$$\mathcal{O}=\ln(\mathcal{A}) = (3/2)[\ln(\mathcal{M}_4)- \ln(\mathcal{M}_2)],$$
where $\mathcal{M}_n$ is defined as
\begin{eqnarray}
 \mathcal{M}_n = \int d^3k ~ k^{n-2} P_X(\vec{k}) W(k R_{\mathcal{G}}),
\end{eqnarray}
$P_X(\vec{k})$ denotes the power spectrum under consideration,
then the variance $\sigma^2_{\mathcal{O}} \equiv {\rm Var}(\mathcal{O})$ is 
simply
\begin{eqnarray}
    \label{eqn:sigA}
 \sigma^2_{\mathcal{O}}  = 
  \bigl(\frac{3}{2} \bigr)^2 \biggl[
   \frac{ {\rm Var}(\mathcal{M}_2) }{\mathcal{M}_2^2}
 + \frac{ {\rm Var}(\mathcal{M}_4) }{\mathcal{M}_4^2}
  - 2 \frac{ {\rm Cov}(\mathcal{M}_2, \mathcal{M}_4) }{\mathcal{M}_2 \mathcal{M}_4}
  \biggr]. \nonumber \\
\end{eqnarray}
Here ${\rm Var}(\mathcal{M}_m) \equiv {\rm Cov}(\mathcal{M}_m, \mathcal{M}_m)$, 
and  ${\rm Cov}(\mathcal{M}_m, \mathcal{M}_n)$ is propagated from $\sigma^2_{P}$
\begin{eqnarray}
 {\rm Cov}(\mathcal{M}_m, \mathcal{M}_n) = 
  \langle \mathcal{M}_m, \mathcal{M}_n \rangle - 
  \langle \mathcal{M}_m \rangle \langle \mathcal{M}_n \rangle &&
   \nonumber \\
 = \frac{2 (2\pi)^2}{V_s} \int dk~d\mu ~ k^{(m+n-2)} 
   W^2(k R_{\mathcal{G}})~  \sigma^2_{P_X}(\vec{k}). ~~&&
\end{eqnarray}
where $\mu$ is the cosine of the angle between $\vec{k}$ and the line of sight,
$V_s$ is the survey volume.
We have assumed the covariance matrix of power spectrum is diagonal, 
\begin{eqnarray}
\label{eqa:covP}                  
 \Cov \bigl( P(\vec{k}_1),~  P(\vec{k}_2)\bigr) 
  =    \delta_D( \veck_1 - \veck_2 ) \frac{2 (2\pi)^2}{V_s} 
  \sigma_{P_X}^2 
\end{eqnarray}

We need to estimate the statistical uncertainty in power spectrum measurements
for the different experiments. For galaxy surveys, the statistical uncertainty 
of $P_{\rm g}(k)$ per Fourier mode 
includes both the cosmic variance and the shot noise due to the finite
number of galaxies:
\begin{eqnarray}
\sigma_{P_{\rm g}} (\vec{k}) = \biggl [
  P_{\rm g} +  \frac{1}{n} \biggr ], 
\end{eqnarray}
As is often done in such estimates (e.g. \cite{SE03}), 
we assume the density of the galaxy $n(z)$ satisfies
$nP(k=0.2)=3$. 

For the 21cm intensity mapping experiment, e.g. the 
cylindrical radio telescope (CRT), in addition to the cosmic variance and 
the shot noise due to the finite number of galaxies, there is also the noise
due to the foreground and receiver \citep{SS09}, so 
$\sigma_{P_{21cm}}(\vec{k})= P_{21cm}(\vec{k})$, with
\begin{eqnarray}
P_{21cm}(\vec{k})&=&  
    p_s^2 \left(P_{HI}(\vec{k}) + \frac{1}{n}\right)+ \nonumber \\
 & &   ~~~ 
\biggl( \frac{k_B (g \bar{T}_{\rm sky} + \bar{T}_a ) \Delta f}{ 
    \sqrt{ t_{\rm int}  \Delta f }} \biggr)^2 V_{R} \nonumber 
\end{eqnarray}
with $p_s = k_B g \bar{T}_{\rm sig} \Delta f$, 
$g=0.8$ is the gain, $t_{\rm int}$ is the integration 
time and $V_R$ is the volume of a pixel,
$\bar{T}_{\rm sig}$ is the average brightness temperature which is estimated as
\begin{eqnarray}
  \bar{T}_{\rm sig} = 188 \frac{x_{HI}(z) \Omega_{H,0} h (1+z)^2 }{H(z)/H_0} \mK,
\end{eqnarray}
with a conservative assumption for the neutral hydrogen fraction 
$x_{HI}(z) \Omega_{H,0}=0.00037$.
$\bar{T}_{\rm sky}$ and $\bar{T}_{\rm a}$ are average sky and antenna noise 
temperatures, which are assumed to equal $10\K$ and $50\K$ respectively.

For the high redshift 21cm experiment such as the MWA, 
we assume that the system temperature of the 
telescope is dominated by 
the sky: $$\bar{T}_{sys}\sim 250[(1+z)/7]^{2.6} \K,$$ 
and the observation time is
$$t_k= (A_e t_{\rm int}/\lambda^2) n(k_{\perp})$$
at $k_{\perp}$,  where $n(k_{\perp})$ is the number of baselines
which observe the transverse component of the 
wavevector. This can be
calculated from the array configuration, here we 
model the antennas distribution of MWA 
as $\rho(r)\sim r^{-2}$, 
with radius $r_m=750$m and a flat core of radius $r_c=20$m.
For the hypothetical follow-up of MWA, denoted as MWA5000,
we have assumed $r_m=2\km$ and  $r_c=80 $m. 

The results of our estimates are given in Sec.4.

\label{lastpage}

\end{document}